\documentclass[useAMS,usenatbib]{mn2e}
\usepackage{graphicx}
\usepackage{enumerate}
\usepackage{epsfig}
\usepackage{comment}

\title[Connection between disc and power-law variability]{The causal connection between disc and power-law variability in hard state black hole X-ray binaries}
\author[P. Uttley et al.]{P. Uttley$^{1}$\thanks{E-mail:
p.uttley@soton.ac.uk}, T. Wilkinson$^{1}$, P. Cassatella$^{1}$, J. Wilms$^{2}$,
K. Pottschmidt$^{3,4}$, \newauthor M. Hanke$^{2}$, M. B\"{o}ck$^{2}$\\
$^{1}$Faculty of Physical and Applied Science, University of Southampton, Southampton SO17 1BJ\\
$^{2}$Dr. Karl Remeis-Observatory and ECAP, University of Erlangen-Nuremberg, Sternwartstrasse 7, 96049 Bamberg, Germany\\
$^{3}$CRESST, University of Maryland Baltimore County, 1000 Hilltop Circle, Baltimore, MD 21250, USA\\
$^{4}$NASA Goddard Spaceflight Center, Astrophysics Science Division, Code 661, Greenbelt, MD 20771, USA}

\begin{document}

\date{}

\pagerange{\pageref{firstpage}--\pageref{lastpage}} \pubyear{2010}

\maketitle

\label{firstpage}

\begin{abstract}
We use the {\it XMM-Newton} EPIC-pn instrument in timing mode to extend spectral time-lag studies of hard state black hole X-ray binaries into the soft X-ray band.  We show that variations of the disc blackbody emission substantially lead variations in the power-law emission, by tenths of a second on variability time-scales of seconds or longer.  The large lags cannot be explained by Compton scattering but are consistent with time-delays due to viscous propagation of mass accretion fluctuations in the disc.  However, on time-scales less than a second the disc lags the power-law variations by a few ms, consistent with the disc variations being dominated by X-ray heating by the power-law, with the short lag corresponding to the light-travel time between the power-law emitting region and the disc.   Our results indicate that instabilities in the accretion disc are responsible for continuum variability on time-scales of seconds or longer and probably also on shorter time-scales. 
\end{abstract}


\section{Introduction}
\label{int}
Rapid X-ray variability on time-scales of seconds to milliseconds is a key characteristic of X-ray binaries (XRBs) which has been studied almost since their discovery.  It was suggested quite early on that the rapid variability observed in XRBs and on much longer time-scales in active galaxies, was produced by instabilities in the `standard' accretion disc  \citep{Lightman74,Shakura76}. Such a model is attractive because the disc can produce variations over a broad range of time-scales, corresponding to the range of unstable disc radii, to match well with the observed variability \citep{Lyubarskii97}.

However, for many years the variability has been associated with the still-mysterious hard X-ray power-law emitting component (e.g. \citealt{Galeev79,Poutanen99,Churazov01,Done07} and references therein).   This is because, although strong variations in the luminosity of the disc blackbody component are seen on long time-scales corresponding to transitions between the soft and hard spectral states, the disc-dominated soft states show very little rapid variability (fractional rms $\sim1$~per cent, e.g. \citealt{Belloni05}).  In contrast, rms of several tens of per cent is seen in the hard spectral states which are energetically dominated by the power-law emission thought to be associated with a hot optically-thin flow, magnetised corona, or even the base of the persistent radio-emitting jets seen in these states (e.g. \citealt{Zdziarski98,Droulans10,Markoff05}).  In addition, the strong QPOs observed during the state-transitions of black hole XRBs (BHXRBs), when both disc and power-law emission are strong, show hard rms-spectra consistent with an origin predominantly in the power-law emitting component \citep{Sobolewska06}. 

In the hard states our understanding of the origin of the variability has been limited by a lack of spectral coverage of the disc, which emits at $kT<0.5$~keV and so is not covered by the bandpass of detectors such as the Proportional Counter Array (PCA) on the {\it Rossi X-ray Timing Explorer}, which is the prime workhorse of X-ray timing studies to date.  Recently we addressed this problem by studying the X-ray variability `covariance spectra'\footnote{The covariance spectrum measures the spectral shape of variations which are correlated with a given reference band, and so is ideal to search for correlated disc and power-law variability, see \citet{Wilkinson09} for details.} of the hard state of the black hole candidate GX~339$-$4 \citep{Wilkinson09} using the soft X-ray response of the EPIC-pn CCD detector on board the {\it XMM-Newton} satellite.  In that work, we found that the disc blackbody emission does vary with an rms amplitude of tens of per cent and that the blackbody variability amplitude is greater relative to the correlated power-law variations on longer time-scales ($>1$~s).  We used these results to argue that, although on time-scales $<1$~s disc blackbody variability could be explained by X-ray heating of the disc by the varying power-law emission, on longer time-scales the disc is intrinsically variable.  A natural physical interpretation is that instabilities in the standard disc are responsible for driving the observed power-law variability on time-scales at least down to a few seconds, perhaps through mass-accretion variations which propagate through the disc \citep{Lyubarskii97,Arevalo06}, before reaching the power-law emitting region. 

An even stronger argument can be made for disc-driven variability if we can establish the {\it causal} relationship between correlated variations in the disc and power-law components, through the detection of X-ray time-lags between variations in different energy bands.  Previous work to measure time-lags in BHXRBs has used proportional counter instruments to measure lags between variations at harder energies, above 2~keV (e.g. \citealt{Miyamoto89,Nowak99}), however we can also do this analysis using the same {\it XMM-Newton} EPIC-pn timing mode data that we use to study the covariance spectrum.  In this Letter we use these data to present the first time-lag study of the hard state which is extended to soft X-rays to cover the disc component.  We confirm that there is indeed a clear signature in the lags at soft energies which shows that variations on time-scales of seconds or longer in the disc blackbody emission lead correlated variations in the power-law emission.  We also show that the sign of the lag changes on time-scales $<1$~s, so that the disc component lags behind the power-law variations by a few milliseconds.  These lags are consistent with light-travel time-lags expected by disc thermal `reverberation', caused by X-ray heating of the disc by the power-law which dominates over the intrinsic disc blackbody variability on these short time-scales.

\section{Observations and Data Reduction}
\label{obs}
We analysed EPIC-pn \citep{Strueder01} timing mode data from the 2004 March 16-19 {\it XMM-Newton} observations of GX-339 in a stable hard state. The data were reduced in the standard manner using {\sc SAS} 10.0.0, processing the raw data products using the {\sc SAS} tools {\sc epsplitproc} and {\sc epfast}, taking account of background flaring and extracting only events with RAWX from columns 31 to 45 using the {\sc SAS} tool {\sc evselect}. Although the tool {\sc epfast} was used on the events file, the Charge Transfer Inefficiency (CTI) corrections were insufficient to remove features at 1.8 and 2.2 keV (Si and Au edges) in the 2004 spectrum. Although the source is relatively bright, previous analyses have shown that the EPIC-pn data are not significantly piled up \citep{Done10}. The {\sc SAS} tools {\sc arfgen} and {\sc rmfgen} were used to generate the ancillary response file (ARF) and redistribution matrix file (RMF) for the 15 columns used for source extraction. Background was not accounted for as it is very small compared to the source and in any case background variations are uncorrelated with those of the chosen reference energy  bands and so do not contribute to the lags or covariance spectra.  The total exposure time for the combined 2004 data was 155~ks.  

\section[]{Analysis and Results}
\subsection{Power and time-lags versus frequency}
\begin{figure}
\hbox{\includegraphics[scale=0.7,angle=-90]{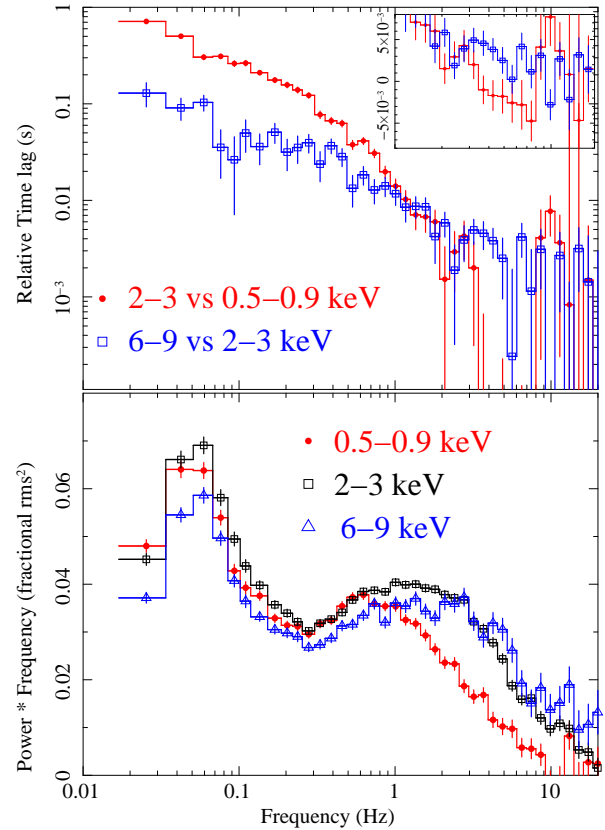}}
\caption{Top panel: GX~339$-$4 2004 hard state lag versus frequency dependence for the medium-soft and hard-medium energy bands.  Positive values correspond to harder photons lagging softer photons. The inset shows the lags on a linear y-axis in the frequency range where the medium-soft lags drop below the hard-medium lags (the x-scale is matched to the scale on the main figure).  Bottom panel: Power spectra for the soft, medium and hard energy bands.}
\label{lagpsd}
\end{figure}
We used the event lists, together with Good Time Intervals (GTIs) excluding telemetry dropouts, to extract light curves in three geometrically-spaced energy bands (soft 0.5--0.9~keV, medium 2--3~keV and hard 6--9~keV)\footnote{There is a factor $\sim3$ separation in average event energy between each band, with values of 750~eV, 2420~eV and 7140~eV for soft, medium and hard bands respectively.} in order to measure the cross-spectra and determine lags between bands which are separated by equal steps in $\log(E)$.  Cross-spectra for pairs of these energy bands were measured in the standard way \citep{Nowak99} for contiguous segments of data of length 59~s (the segment-length necessitated excluding data containing short gaps due to telemetry drop-outs).  The cross-spectra were averaged over segments and in geometrically spaced frequency bins with frequency separation $\nu\rightarrow1.15\nu$.  We measured phase lags from the argument of the average cross-spectrum in each bin and divided by $2\pi$ times the average frequency of the bin to give the time-lag.  The resulting lag-frequency spectra are shown in Fig.~\ref{lagpsd}, together with the power spectra of the measured energy bands.

From the lag-frequency spectra, we note two striking differences between the lag behaviour in the soft band and the behaviour at harder energies which has already been well-studied by {\it RXTE}.  First, although we chose the energy bands to be separated by roughly-equal steps in $\log(E)$, at low frequencies the medium-soft lags are a factor $\sim6$ times larger than the hard-medium lags.  Therefore, by extending the measurement of lags into the soft band we have uncovered a significant increase in the lags compared with the values expected from extrapolation of the log-linear scaling with energy observed by {\it RXTE} above 3~keV (e.g. \citealt{Nowak99}), which would give similar lags between pairs of bands separated by the same $\log(E)$.  It is also interesting to note that although at low frequencies the medium-soft and hard-medium lag-frequency relations show similar power-law shapes (time-lag $\tau \propto \nu^{-0.7}$, as previously observed), at frequencies approaching 1~Hz the medium-soft lags start to drop rapidly, to the point (above 3~Hz) where they drop below the hard-medium lags.  This drop in medium-soft lags seems to correspond roughly to a drop in the soft band power relative to other bands at frequencies approaching and above $\sim1$~Hz.  In \citet{Wilkinson09} we used the covariance spectra measured on different time-scales to associate this drop in soft band power at high frequencies with a drop in the amplitude of variability intrinsic to the blackbody-emitting disc, so we might expect the disc to play a similar role in explaining the medium-soft lag behaviour. 

\subsection{Lag versus energy and covariance spectra}
To better understand the origin of the lag behaviour associated with the soft band, we carried out a Fourier-frequency-resolved study of lag versus energy.  To optimise signal-to-noise we measured the cross-spectra for many narrow energy bins with respect to a fixed broad reference band covering the range 0.54--10.08~keV (the reference energies are selected to match the start and end of energy bins).  In order to remove the correlated Poisson noise part of the variability from the cross-spectrum we must remove the energy bin light curve from the reference band light curve before measuring the cross-spectrum.  Strictly-speaking this means that the reference band is slightly different for each energy bin, but due to the narrow bin widths, the effects on the relative lags are negligible compared to the errors (see also the discussion by \citealt{Zoghbi10}).
\begin{figure*}
\hspace{1.5cm}
\hbox{\includegraphics[scale=0.55,angle=-90]{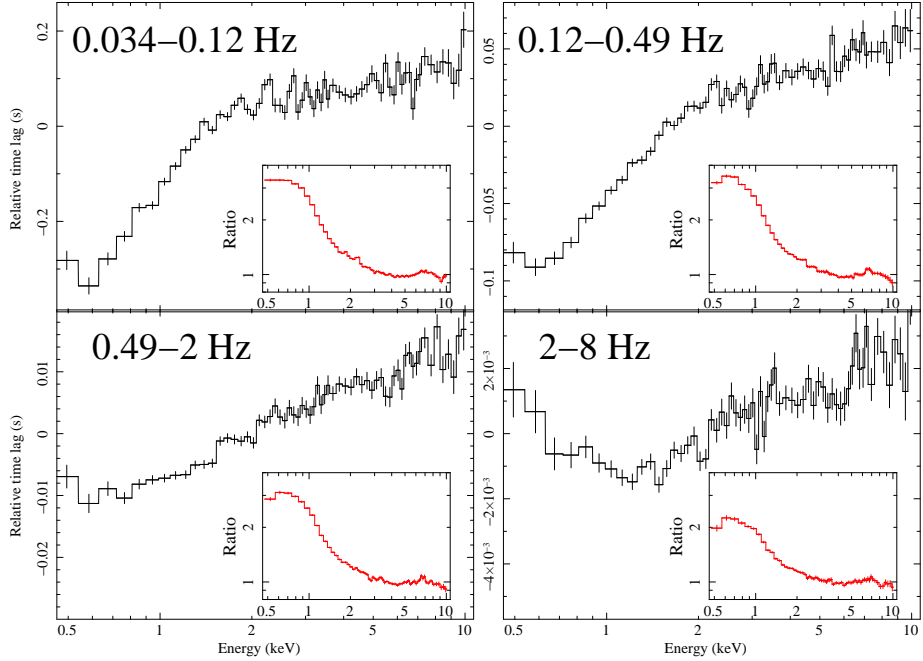}}
\caption{Lag-energy spectra for four temporal frequency ranges.  The lags are all measured relative to the same 0.54-10.08~keV reference band and plotted so that energies with more-positive values of lag are lagging behind energies with less-positive values of lag.  The insets show the covariance spectra for the same energy range and frequencies, plotted as a ratio to a power-law with photon index fixed at $\Gamma=1.55$, absorbed by a $6\times10^{21}$~cm$^{-2}$ neutral Galactic column and fitted to the 3--10~keV range.  Note that the scales for the covariance plots are the same, showing that the switch in soft lag behaviour at high frequencies corresponds to a drop in amplitude of disc variability relative to the power-law variability.}
\label{specprod}
\end{figure*}

After measuring the cross-spectrum for each channel with respect to the reference band, we can determine the relative lag versus energy for a selected frequency-range by averaging the cross-spectrum over that range (e.g. see \citealt{Nowak99,Kotov01}).  We can use the frequency-averaged cross-spectra to determine phase lags and hence time-lags in the usual way.  We can also use the averaged cross-spectrum to make frequency-resolved covariance spectra, which can be obtained by dividing the square-root of the modulus-squared of the cross-spectrum by the rms of the reference band, in a manner analogous to the way the time-domain covariance spectrum is measured using the cross-correlation function \citep{Wilkinson09}.

Fig.~\ref{specprod} shows the frequency-resolved lag-energy spectra for four frequency ranges, each corresponding to an approximately factor~4 increase in frequency starting with 0.034~Hz and ending at 8~Hz.  Covariance spectra for the same frequency ranges are shown as insets and are plotted as ratio plots with respect to a simple absorbed power-law model (see Figure caption for more details). 

The frequency-resolved lag-energy spectra show clearly that the soft excess emission associated with the disc blackbody is responsible for the complex soft lag behaviour seen in Fig.~\ref{lagpsd}.  The increased amplitude of the low-frequency medium-soft lags results from a significant downturn in the lag-energy relation, away from the log-linear law which applies above 2~keV.  The covariance spectra show that this downturn is associated with energies where variable disc blackbody emission starts to become significant.  Therefore, the disc blackbody variations are significantly {\it leading} the variations of the power-law emission which dominate at harder energies.

The covariance spectra show that at high frequencies the disc blackbody variability is still present but drops in amplitude.  The same result was found in the time-domain covariance spectra \citep{Wilkinson09}, our interpretation being that at these frequencies, intrinsic disc variability is small compared to blackbody variations driven by X-ray heating of the disc by the power-law emission.  This interpretation appears to be confirmed by the lag-energy spectra.  The downturn in low-frequency lag-energy spectra becomes an upturn at high frequencies, so that now the soft photons lag the medium-energy photons (which explains the sharp change in medium-soft lags seen above 1~Hz in Fig.~\ref{lagpsd}). This behaviour can be simply explained if the mechanism for disc blackbody variability is switching from being dominated by variable viscous heating intrinsic to the disc on time-scales below 1~s, to being dominated by X-ray heating by the illuminating power-law on shorter time-scales. The high-frequency soft lags, which correspond to tens of $R_{\rm G}$ light-crossing time, can be associated with the light travel time from the power-law emitting regions to the blackbody-emitting regions of the disc.

\subsection{Lags in other hard state systems}
\begin{figure}
\hbox{\includegraphics[scale=0.33,angle=-90]{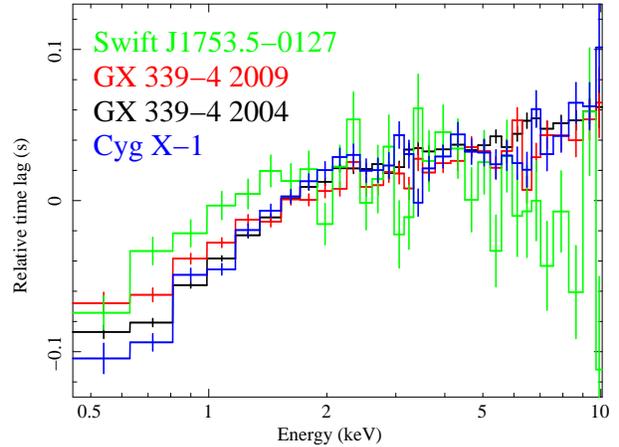}}
\caption{Lag-energy spectra of 0.125--0.5~Hz variations for several hard state systems.}
\label{lagcompare}
\end{figure}

To show that other hard state systems show the same low-frequency lag behaviour as GX~339$-$4, we plot in Fig.~\ref{lagcompare} lag versus energy for the 0.125-0.5~Hz range for EPIC-pn timing mode data from {\it XMM-Newton} observations of hard states in GX~339$-$4, Swift~J1753.5$-$0127 and Cyg~X-1 as well as for GX~339$-$4 in a lower-flux state than observed in 2004\footnote{Swift~J1753.5$-$0127 was observed on 2006 March 24, GX~339$-$4 and Cyg~X-1 were observed on 2009 March 26 and 2009 December 2 respectively.}.  The lags at higher frequencies cannot be compared with the GX~339$-$4 2004 data due to larger error bars caused by the shorter exposure times of these observations (40~ks, 16~ks and 31~ks for Swift~J1753.5$-$0127, Cyg~X-1 and GX~339$-$4 2009 respectively).  However, in the 0.125-0.5~Hz range the other systems, as well as the GX~339$-$4 2009 data, show a downturn in lag versus energy in soft X-rays which is similar to that seen in the GX~339$-$4 2004 data.
\section{Discussion}
\label{discussion}
By extending the measurement of energy-dependent lags to soft energies, we have shown that on time-scales of seconds or longer, disc blackbody variations lead variations in the power-law component.  The natural implication of this result is that the disc variations drive the power-law variations, so we now consider the physical origin of this causal connection.  The medium-soft lags of 0.1~s or longer seen in variations on time-scales of seconds correspond to light-travel times of thousands of gravitational radii.  Coronal Comptonisation models have been invoked to explain the log-linear form of the lags seen at harder energies \citep{Payne80,Kazanas99}, but they struggle to explain even the more moderate hard-medium lags seen at low frequencies without invoking an unfeasibly large Comptonising region \citep{Nowak99}.  A recent more energetically-feasible model invokes Comptonisation by the persistent hard state jet to produce the hard lags \citep{Reig03,Kylafis08}.  In this model, seed photons are fed into the base of the jet (e.g. from the disc), and the anisotropy of Compton scattering along the jet axis means that upscattered photons can traverse large distances in the jet before escaping to the observer, to produce large lags relative to the photons undergoing fewer upscatterings.   However, in such a model we would expect the medium-soft lags to be reduced relative to the hard-medium lags, not increased as observed here, since the direct seed photons from the disc at soft energies will dilute the log-linear lags introduced by the upscattered continuum. The medium-soft lag would only increase relative to the hard-medium lag if the medium-soft lag represents the actual travel time from the seed photon source to the scattering electron population, i.e. the seed photons must first interact with the jet not at its base but thousands of gravitational radii above the disc.  This geometry raises significant problems for explaining the energetics and variability of the emission (since the jet at that height would subtend only a small solid angle as seen from the disc), not to mention the shape and strength of the iron $K\alpha$ line, which due to geometric and beaming effects would be much narrower and weaker than observed, if the disc were illuminated by the jet from such a great height (e.g. \citealt{Markoff04}).

A likely explanation of the large medium-soft lags observed on time-scales of seconds is that the lags are associated with the generation of and propagation of accretion fluctuations within the inner regions of the disc before they reach the corona \citep{Lyubarskii97,Kotov01,Arevalo06}.  In fact, the observed lag behaviour on these and sub-second time-scales is exactly that expected from our earlier disc-variability interpretation of the covariance spectrum \citep{Wilkinson09}.  In the present work, we have used causal information to greatly strengthen that interpretation, so that accretion instabilities can be firmly identified as the source of X-ray variability, at least on time-scales of seconds.  Furthermore, the unstable accretion flow is now shown to be the standard disc, with all the attendant physical implications (see discussion in \citealt{Wilkinson09}), and not a hot optically thin flow. 

The disc propagation model can explain the medium-soft lags, however the hard-medium lags - produced where the disc blackbody does not contribute to the X-ray spectrum - still require explanation.  These lags may still be related to Compton upscattering of seed photons as the accretion fluctuations reach the base of the jet, e.g. in a hybrid model of disc propagation and the model of \citet{Reig03}.  However, given the common $\nu^{-0.7}$ frequency-dependence of the low-frequency lags which seems to be independent of the energy-bands chosen, it seems likely that the lags at hard energies are directly related to the same underlying propagation mechanism which produces the medium-soft lags.  For example, lags at harder energies can be simply produced if the mass-accretion fluctuations in the disc preferentially generate softer and then harder power-law emission as they propagate inwards.  Such effects could be produced if there is a soft coronal component above the disc while the hard emission is produced centrally (e.g. \citealt{Kotov01,Arevalo06}).

The switch in lag behaviour on short time-scales suggests that the disc lags the power-law variations by a few ms on these time-scales, consistent with the light-travel lags expected from a power-law component separated from the disc by only tens of gravitational radii at most. This lag signature represents the first evidence in BHXRBs for a disc thermal reverberation lag.  This is the time-lag due to the light travel-time from the central power-law emitting region to the disc where the hard X-rays are reprocessed into blackbody emission.  In order to use these lags to map the disc, e.g. to constrain the inner radius, we would need to make assumptions about the geometry of the power-law emitting region and the emissivity profile of the disc blackbody.  This modelling effort is beyond the scope of this work.

Finally, it is important to note that although the lags observed for variations on time-scales of less than a second appear to be caused by X-ray heating of the disc, this does not necessarily imply that the variability on these shorter time-scales is not also generated in the disc.  An accretion instability at small disc radii will modulate only disc emission inside that radius (e.g. \citealt{Arevalo06}), so the intrinsic variability in disc emission will be small and likely to be dominated by X-ray heating effects when the mass fluctuations reach the central power-law emitting region.

\section*{Acknowledgments}
We would like to thank the anonymous referee for valuable comments.  PU is supported by an STFC Advanced Fellowship and TW is supported by an STFC postgraduate studentship grant. The research leading to these results has received funding from the European Community's Seventh Framework Programme (FP7/2007-2013) under grant agreement number ITN 215212 ``Black Hole Universe''. This work was partly funded by the Bundesministerium f\"ur Wirtschaft and Technologie through Deutsches Zentrum f\"ur Luft-
und Raumfahrt grants 50 OR 0701 and 50 OR 0808. This work is based on observations obtained with {\it XMM-Newton}, an ESA science mission with instruments and contributions directly funded by
ESA Member States and NASA.

\label{lastpage}

\end{document}